\documentclass{article}
\usepackage{times}
\usepackage[latin1]{inputenc}

\usepackage{pslatex}
\usepackage{amsmath}
\usepackage{amssymb}
\usepackage{amsfonts}
\usepackage{amsthm}
\usepackage{hyperref}
\usepackage{float}
\floatstyle{ruled}
\restylefloat{table}
\restylefloat{figure}

\usepackage{color}
\newcommand{\p}{{\sf P}}
\newcommand{\cJ}{{\cal J}}
\newcommand{\gO}{\Delta\Delta}

\newcommand{\filterscheme}{{\cal S}}
\newtheorem{fact}{Fact}[section]

\newtheorem{definition}[fact]{Definition}
\newtheorem{theorem}[fact]{Theorem}
\newcommand{\Ruled}[2]{\mbox{$\displaystyle\frac{#1}{#2}$}}
\newcommand{\md}{}

\newcommand{\forget}[1]{}

\newcommand{\set}[1]{\{#1\}}
\newlength{\lrulenamec}
\newcommand{\ruled}[2]{\mbox{$\displaystyle\frac{#1}{#2}$}}

\newcommand{\into} {\cap}
\newcommand{\ti}{\lambda\into^{\tytv}}

\newcommand{\con}{{{\sf C}\!\!\!\!{\sf C}}}

\newcommand{\tyt} {\Omega}

\newcommand{\conv}{\con^{\tytv}}

\newcommand{\tva} {A}
\newcommand{\tvb} {B}
\newcommand{\tvc} {C}

\newcommand{\tvd} {D}

\newcommand{\sar}[1]{\Sigma^{#1}}
\newcommand{\tleq} {\leq}
\newcommand{\tleqt}{\tleq_{\tytv}}

\newcommand{\binto} {\bigcap}
\newcommand{\teqt}{{}{\sim_{\tytv}}}
\newcommand{\notteqt}{{}{\mbox{$\not\sim$}_{\tytv}}}

\newcommand{\tytv} {\bigtriangledown}
\newcommand{\calT} {\sar{\tytv}}

\newcommand{\ctas}{\vdash^\tytv}

\newcommand{\ax} {\mbox{Ax}}
\newcommand{\axtyt} {\mbox{Ax-$\tyt$}}

\newcommand{\arrI} {\arr  \mbox{\rm I}}

\newcommand{\intI} {\into \mbox{\rm I}}
\newcommand{\arE} {\arr  \mbox{\rm E}}

\newcommand{\FX} {X}
\newcommand{\FY} {Y}
\newcommand{\SF}{{{\cal F}^\tytv}}
\newcommand{\FFF}[1]{\uparrow^{#1}}

\newcommand{\G}{\Gamma}

\newcommand{\en} {\rho}
\newcommand{\ag} {\models}
\newcommand{\dlsqb}{[\![}
\newcommand{\drsqb}{]\!]}
\newcommand{\interpretation}[3]{\dlsqb{#1}\drsqb^{#3}_{#2}}

\newcommand{\three}[3]{\langle {#1},{#2},{#3} \rangle}

\newcommand\type{{\sf T\!\!\!\!\!T}}
\newcommand{\arr}{\rightarrow}

\newcommand{\semic}{\! : \!}

\newcommand{\naturali}{\mbox{\rm I$\!$N}}

\newcommand{\ctastas}{\ctas}

\newcommand{\FV}{\mbox{FV}}
\newcommand{\typetytv}{\type^{\tytv}}
\newcommand{\tM}{{\sf t}}
\newcommand{\tN}{{\sf u}}
\newcommand{\tE}{{\sf e}}

\begin{document}

\title{Intersection Types and Lambda Theories
\thanks{Partially supported by MURST Cofin'00 AITCFA  Project, MURST
Cofin'01 COMETA Project, and by  EU within the FET - Global Computing initiative, project DART IST-2001-33477.} }
\author{
M.Dezani-Ciancaglini\thanks{Dipartimento di Informatica, Universit\`a di Torino, Corso
Svizzera 185, 10149 Torino, Italy  {\tt dezani@di.unito.it}}
\and S.Lusin
\thanks{Dipartimento di Informatica, Universit\`a di Venezia,
{via Torino 153, 30170 Venezia, Italy}  {\tt slusin@oink.dsi.unive.it}}}
\date{}
\maketitle

\begin{abstract}
We illustrate the use of intersection types as a semantic tool for
showing properties of the lattice of $\lambda$-theories.  Relying
on the notion of {\em easy intersection type theory} we
successfully build a filter model in which  the
interpretation of an arbitrary simple easy term is any filter which can be described in an uniform way by a predicate. This allows us to prove the consistency of a
well-know $\lambda$-theory: this consistency has interesting consequences on the algebraic structure of the lattice of $\lambda$-theories.
\end{abstract}

\section*{Introduction}

Intersection types were introduced in the late 70's by Dezani and
Coppo \cite{coppdeza80,coppdezavenn80,barecoppdeza83}, to overcome
the limitations of Curry's type discipline. They are a very
expressive type language which allows to describe and capture
various properties of $\lambda$-terms. For instance, they have
been used in \cite{pott80}
 to give the first type theoretic characterization of {\em strongly
 normalizable} terms and in \cite{coppdezazacc87}  to
 capture {\em persistently normalizing terms} $\md$ and {\em
 normalizing terms}$\md$. See \cite{dezahonsmoto2000} for a more
complete account of this line of research.

Intersection types have a very significant realizability semantics
with respect to applicative structures. This is a generalization
of Scott's natural semantics \cite{scot72} of simple types.
According to this interpretation types denote subsets of the
applicative structure, an arrow type $A\to B$ denotes the sets of
points which map all points belonging to the interpretation of $A$
to points belonging to the interpretation of $B$, and an
intersection type $A\into B$ denotes the intersections of the
interpretation of $A$ and the interpretation of $B$. Building on
this, intersection types have been used in \cite{barecoppdeza83}
to give a proof of the completeness of the natural semantics of
Curry's simple type assignment system in applicative structures,
introduced in \cite{scot72}.

Intersection types have also an alternative semantics based on
{\em duality} which is related to Abramsky's {\em Domain Theory in
Logical Form } \cite{abra91}. Actually it amounts to the
application of that  paradigm to the special case of
$\omega$-algebraic complete lattice models of pure
$\lambda$-calculus, \cite{coppdezahonslong84}.  Namely, types correspond to
{\em compact elements}: the type $\Omega$ denoting  the least
element, intersections denoting {\em joins}  of compact elements,
and arrow types denoting {\em step functions} of compact elements.
A typing judgment then can be interpreted as saying that a given
term belongs to a pointed compact open set in a $\omega$-algebraic
complete lattice model of $\lambda$-calculus. By duality, type
theories give rise to {\em filter $\lambda$-models}. Intersection
type assignment systems can then be viewed as {\em finitary
logical} descriptions of the interpretation of $\lambda$-terms in
such models, where the meaning of a $\lambda$-term is the set of
types which are deducible for it.
This duality lies at the heart of the success of intersection
types as a powerful tool for the analysis of $\lambda$-models, see
\cite{plot93} and the references there.

A key observation is that the $\lambda$-models we build out of intersection types
differ only for the {\em preorder relations} between types. Changing
these preorders in fact allow us to give different
interpretations to $\lambda$-terms. In all these preorders crucial are the
equivalences between atomic types and intersections of arrow types: therefore
{\em type isomorphisms} are the corner stones of filter model constructions.

In \cite{aleslusi02} Alessi and Lusin faced the issue of easiness
proofs of $\lambda$-terms from the semantic point of view (we
recall that a closed term $\tE$ is {\em easy} if, for any other
closed term $\tM$, the theory $\lambda\beta + \{\tM=\tE\}$ is
consistent). Actually the mainstream of easiness proofs is based
on the use of syntactic tools (see \cite{kupe97} and the
references there). Instead, very little literature can be found on
easiness issues handled by semantic tools, we can mention the
papers \cite{zylb91}, \cite{baetboer79},  \cite{alesdezahons01},
\cite{aleslusi02}.

Going in the direction of \cite{alesdezahons01}, in
\cite{aleslusi02} Alessi and Lusin introduced the notion of {\em
simple easiness}: roughly speaking, an unsolvable term $\tE$ is
simple easy if, for each filter model ${\cal F}^{\tytv}$ built on
an easy intersection type theory $\Sigma^{\tytv}$, any type $Z$ in
$\Sigma^{\tytv}$, we can expand $\Sigma^{\tytv}$ to a new easy
intersection type theory $\Sigma^{\tytv'}$ such that the
interpretation of $\tE$ in ${\cal F}^{\tytv'}$ is the sup of the old
interpretation of $\tE$ in ${\cal F}^{\tytv}$ and of the filter
generated by $Z$.

A consequence is that simple easiness is a stronger notion than
easiness. A simple easy term $\tE$ is easy, since, given an
arbitrary closed term $\tM$, it is possible to build (in a canonical
way) a non-trivial filter model which equates the interpretation
of $\tE$ and $\tM$.

Besides of that, simple easiness is interesting in itself, since
it has to do with minimal sets of axioms which are needed in order
to give the easy terms certain types. This can be put at work to interpret
easy terms by filters which can be described in an uniform way by predicates.

Building on the duality between type intersections and joins, arrows and step functions, given an arbitrary simple easy term we build a $\lambda$-model in which this term is interpreted as the join. 
In this way we can prove the consistency of an interesting $\lambda$-theory. This consistency has been used in \cite{lusisali02} to show that there exists a sublattice of the lattice of $\lambda$-theories
which satisfies a restricted form of distributivity, called meet
semidistributivity,
and a nontrivial congruence identity (i.e., an identity in the language
of lattices enriched by the relative product of binary relations).

The present paper is organized as follows. In
Section~\ref{theories} we present easy intersection type theories
and type assignment systems for them. In
Section~\ref{filter-models} we introduce $\lambda$-models based on
spaces of filters in easy intersection type theories. Section
\ref{section-simple-easy-terms} gives the main contribution of the present paper: each simple easy term can be interpreted as an arbitrary filter which can be described in an uniform way by a predicate. Finally in Section \ref{appl} we apply our result to show the consistency of a $\lambda$-theory which has interesting consequences on the algebraic properties of the lattice of $\lambda$-theories.

\section{Intersection Type Assignment
Systems}\label{intersecion-type-intro} \label{theories} {\em
Intersection types} are syntactical objects built inductively by
closing a given set $\con$ of {\em type atoms} (constants) which
contains the universal type $\tyt$ under the {\em function type}
constructor $\arr$ and the {\em intersection} type constructor
$\into$.
\begin{definition}[Intersection Type Language]$\;$\\
Let $\con$ be a countable set of constants such that
$\tyt\in\con$. The {\em intersection type language} over $\con$,
denoted by
  $\type=\type(\con)$ is defined by the following
abstract syntax:
 \[
\type = \con \mid \type \arr  \type \mid \type \into \type.
\]
\end{definition}
Notice that the most general form of an intersection type is a
finite intersection of arrow types and type constants.
\smallskip

\noindent
{\bf Notation} Upper case Roman letters i.e.\
$\tva,\tvb,\ldots$, will denote
  arbitrary types. Greek letters will denote constants in $\con$.
    When writing intersection types we shall use the
  following convention: the constructor $\into$ takes precedence over
  the constructor $\arr $ and it associates to the right. \smallskip

\noindent Much of the expressive power of intersection type
disciplines comes from the fact that types can be endowed with a
{\em preorder relation} $\leq$, which induces the structure of a
meet semi-lattice with respect to $\into$, the top element being
$\tyt$. We recall here the notion of {\em easy\/} intersection type
theory as first introduced in \cite{aleslusi02}.

\begin{definition}[Easy intersection type
theories]\label{intersection-type-theories}$\;$\\ Let
$\type=\type(\con)$ be an intersection type language. The {\em
easy intersection type theory\/} ({\em eitt} for short)
$\Sigma(\con,{\tytv})$ over $\type$ is the set of all judgments
$\tva\tleq\tvb$ derivable from $\tytv$, where $\tytv$ is a
collection of axioms and rules such that (we write $A\sim B$ for
$A\tleq B \;\&\; B\tleq A$):
\begin{enumerate}
\item $\tytv$ contains the
 set $\overline{\tytv}$ of axioms and rules:
 $$\begin{array}[b]{rlrl}
 \mbox{(refl)}&\tva\tleq\tva
 &\mbox{(idem)}&\tva\tleq\tva\into\tva\\[0.5em]
\mbox{(incl$_L$)}& \tva \into \tvb \tleq \tva &\mbox{(incl$_R$)}&
\tva \into \tvb \tleq \tvb\\[0.5em] \mbox{(mon)}&\ruled{ \tva
\tleq \tva'\;\;\; \tvb \tleq \tvb'}{
                     \tva \into \tvb \tleq \tva' \into \tvb'}
&\mbox{(trans)}&\ruled{ \tva \tleq \tvb\;\;\; \tvb \tleq \tvc}{
                     \tva \tleq \tvc}\\[1em]
\mbox{($\tyt$)}&\tva\tleq\tyt &\mbox{($\tyt$-$\eta$)} &
\tyt\tleq\tyt\to\tyt\\[0.5em] \mbox{($\to$-$\into$)} & (A\to
B)\into(A\to C)\tleq A\to B\into C &\mbox{($\eta$)} &\ruled{
A'\tleq A\;\;\; B\tleq B'}{ A\to B \tleq A'\to B'}
\end{array}$$
\item
further axioms can be of the following two shapes only:
\[\begin{array}{l}
\psi\tleq\psi',\\ \psi\sim\binto_{h\in H} (\xi_{h}\rightarrow
E_{h}).
\end{array}\]
where $\psi,\psi',\xi_{h}\in \con$, $E_{h}\in\type$, and
 $\psi,\psi'\not\equiv\tyt$;
 \item $\tytv$ does not contain further rules;
\item
for each $\psi\not\equiv\tyt$ there is exactly one axiom in
$\tytv$ of the shape $\psi\sim A$;
\item
let $\tytv$ contain $\psi\sim \binto_{h\in H} (\xi_{h}\rightarrow
E_{h})$ and $\psi'\sim \binto_{k\in K} (\xi'_{k}\rightarrow
{E'}_{k})$. Then $\tytv$ contains also $\psi\tleq\psi'$ iff for
each $k\in K$, there exists $h_{k}\in H$ such that ${\xi'}_{k}\leq
\xi_{h_{k}}$ and $E_{h_{k}}\leq {E'}_{k}$ are both in $\tytv$.
%$\;$
\end{enumerate}
\end{definition}
%\vspace{.2in}
 \noindent Notice that:\\ (a) since
$\tyt\sim\tyt\to\tyt\in\Sigma(\con,{\tytv})$ by  ($\tyt$) and
($\tyt$-$\eta$), it follows that all atoms in $\con$  are
equivalent to suitable (intersections of) arrow types;\\  (b)
$\into$ (modulo $\sim$) is associative and commutative;\\ (c) in
the last clause of the above definition $E'_k$ and $E_{h_k}$ must
be constant types for each $k\in K$.
\smallskip

\noindent
{\bf Notation} When we consider an eitt
$\Sigma(\con,{\tytv})$, we will write $\conv$ for $\con$,
$\type^{\tytv}$ for $\type(\con)$ and $\sar{\tytv}$ for
$\Sigma(\con,\tytv)$. Moreover  $\tva\tleqt\tvb$ will be short for
  $(\tva\tleq\tvb)\in\sar{\tytv}$ and $A\teqt B$ for $
  A\tleqt B\tleqt A$.
  We will consider syntactic equivalence ``$\equiv $''  of types
  up to associativity
and commutativity of $\into$. \\

A nice feature of eitts is that the order between intersections of arrows  agrees with the order between joins of step functions. This is stated in the following theorem, whose proof can be found in \cite{aleslusi02}.

\begin{theorem}\label{strong-beta-are-beta}$\;$\\
For all $I$, and $\tva_i, \tvb_i, \tvc, \tvd \in \type^{\tytv}$,
\[
\binto_{i \in I}(\tva_i \arr  \tvb_i)\tleqt \tvc \arr  \tvd
\;\Rightarrow\;\exists J \subseteq I. \tvc \tleqt \binto_{i \in
J}\tva_i\;\&\;\binto_{i \in J}\tvb_i\tleqt  \tvd,\] provided that
$\tvd\notteqt\tyt$.
    \end{theorem}

Before giving the crucial notion of {\em intersection-type
assignment system}, we introduce bases and some related
definitions.
\begin{definition}[Bases]\label{definition-for-type-assignment}$\;$
\begin{enumerate}
\item
 A ${\tytv}$-{\em basis} is a (possibly infinite) set
  of statements of the shape $x\semic\tva$, where $\tva\in\type^{\tytv}$,
  with all variables distinct.
  \item
  If $\Gamma$ is a ${\tytv}$-{basis}
  and $\tva\in\type^{\tytv}$ then
 $\Gamma,x\semic A$ is short for $\Gamma\cup\{x\semic A\}$ when $x\notin\Gamma$.
\end{enumerate}
\end{definition}
\begin{definition}[The type assignment system]\label{typeass}$\;$\\
   The {\em intersection type
    assignment system}\ relative to the eitt $\sar{\tytv}$, notation
  ${\ti}$, is a formal system for deriving judgements of the
  form $\Gamma \ctas \tM:\tva$, where
  the {\em subject\/} $\tM$ is an untyped $\lambda$-term,
  the {\em predicate\/} $\tva$ is  in $\type^{\tytv}$,
  and $\Gamma$ is a ${\tytv}$-basis.
Its axioms and rules are the following:
\[\begin{array}{rlrl}
 (\ax) & \Ruled{(x\semic\tva)\in\Gamma}{\Gamma \ctas x\semic\tva} &(\axtyt) &  \Gamma \ctas
\tM:\tyt\\[1em]
 (\arrI ) & \Ruled{\Gamma, x\semic\tva\ctas \tM:\tvb}{\Gamma \ctas \lambda x.\tM:\tva\arr
\tvb} & (\arE) & \Ruled{\Gamma \ctas
\tM:\tva\rightarrow\tvb\;\;\Gamma \ctas \tN:\tva}{\Gamma \ctas
\tM\tN:\tvb} \\[1em] (\intI) & \Ruled{\Gamma \ctas \tM:\tva\;\;\Gamma
\ctas \tM:\tvb}{\Gamma \ctas \tM:\tva\into\tvb} & (\tleqt) &  \Ruled
{\Gamma \ctas \tM:\tva\;\;\tva\tleqt\tvb}{\Gamma \ctas \tM:\tvb}
\end{array}\]
\end{definition}
As usual we consider $\lambda$-terms modulo $\alpha$-conversion.
Notice that intersection elimination rules $$ (\into E)\ \
\ruled{\G \ctas \tM:\tva\into\tvb}{\G \ctas \tM:\tva} \ \ \ \
\ruled{\G \ctas \tM:\tva\into\tvb}{\G \ctas \tM:\tvb}$$ can be
immediately proved to be derivable in all ${\ti}$.\bigskip

We end this section by stating a Generation Theorem (proved in \cite{aleslusi02}).
\begin{theorem}[Generation Theorem]\label{gen-l}$\;$
\begin{enumerate}
\item \label{gen-l1} Assume $\tva\notteqt\tyt$.
$\G\ctas x:\tva$ iff $(x\semic\tvb)\in \G$ and $\tvb\tleqt \tva$
for some $\tvb \in \type^{\tytv}$.
\item \label{gen-l4}
$\G\ctas \tM\tN:\tva$ iff $\G\ctas \tM:\tvb\arr \tva$, and $\G\ctas
\tN:\tvb$ for some $\tvb \in \type^{\tytv}$.
\item \label{gen-l5}
$\G\ctas \lambda x.\tM:\tva$ iff $\G, x\semic\tvb_i\ctas \tM:\tvc_i$
and $\binto_{i \in I}(\tvb_i\arr \tvc_i)\tleqt \tva$, for some $I$
and $B_i, C_i\in \type^\tytv$.
\item \label{gen-l6}
$\G\ctas \lambda x.\tM:\tvb \arr \tvc$ iff $\G, x\semic\tvb\ctas
\tM:\tvc$.
%$\;$
\end{enumerate}
\end{theorem}

\section{Filter Models}\label{filter-models}
In this section we discuss how to  build $\lambda$-models out of
type theories. We start with the definition of {\em filter\/} for
eitt's. Then we show how to turn the space of filters into an
applicative structure. Finally we will define a notion of
interpretation of $\lambda$-terms and state that we get
$\lambda$-models ({\em filter models}).

Filter models arise naturally in the context of those
generalizations of Stone duality that are used in representing
domain theory in logical form (see \cite{abra91},
\cite{coppdezahonslong84}). This approach provides
a conceptually independent semantics to in\-ter\-sec\-tion
ty\-pes, the {\em lattice semantics}. Types are viewed as {\em
compact elements} of domains. The type $\Omega$ denotes the least
element, intersections denote joins of compact elements, and arrow
types allow to internalize the space of continuous endomorphisms.
Following the paradigm of Stone duality, type theories give rise
to filter models, where the interpretation of $\lambda$-terms can
be given through a finitary logical description.

\begin{definition}\label{definition-of-filters}$\;$
 \begin{enumerate}
\item\label{filter1}
 A $\tytv$-filter (or a filter over $\typetytv$) is a set
$\FX\subseteq \typetytv$ such that:
\begin{itemize}
\item
$\tyt \in \FX$;
\item
if $\tva \tleqt \tvb$ and $\tva \in \FX$, then $\tvb \in \FX$;
\item
if $\tva,\tvb \in \FX$, then $\tva\into\tvb \in \FX$;
\end{itemize}
\item\label{filter2} $\SF$ denotes the set of $\tytv$-filters over
$\typetytv$;
\item\label{filter3} if $\FX\subseteq \typetytv$,
$\FFF{\tytv}\FX$ denotes the $\tytv$-filter generated by $\FX$;
\item\label{filter4} a $\tytv$-filter is {\em principal} if it is of the shape $\FFF{\tytv}
  \{ \tva\}$, for some type $\tva$. We shall denote $\FFF{\tytv} \{ \tva\}$
  simply by $\FFF{\tytv} \tva$.
%$\;$
\end{enumerate}
\end{definition}

It is well known that $\SF$ is an $\omega$-algebraic cpo, whose
compact (or finite) elements are the filters of the form $\FFF{\tytv}
A$ for some type $A$ and whose bottom element is $\FFF{\tytv}
\tyt$.\medskip

Next we endow the space of filters with the notions of application
and of $\lambda$-term interpretation. Let ${\sf Env}_\SF$ be the
set of all mappings from the set of term variables to $\SF$.

\begin{definition}\label{application}$\;$
\begin{enumerate}
\item \label{application1}
Application $\cdot:\SF\times\SF\rightarrow\SF$  is defined as
\[ X\cdot Y =\{ \tvb\mid \exists \tva\in \FY.\tva\to\tvb\in
\FX\}.\]
\item \label{application2}
The interpretation function: $\interpretation{\;}{\;}{\tytv}:
\Lambda \times {\sf Env}_\SF \to \SF$ is defined by
\[\interpretation{\tM}{\rho}{\tytv} =
\{\tva \in \typetytv \mid \exists \Gamma\ag\en.\; \Gamma \ctastas
\tM:\tva\},\] where $\rho$ ranges over ${\sf Env}_\SF$ and
$\Gamma\ag\en$ if and only $(x\semic \tvb)\in \Gamma$ implies
$\tvb\in\en(x)$.
\item
The triple $\three{\SF}{\cdot}{\interpretation{\;}{\;}{\tytv}}$ is
called the {\em filter model} over $\calT$.
%$\;$
\end{enumerate}
\end{definition}

Notice that previous definition is sound, since it is easy to
verify that $X\cdot Y$ is a $\tytv$-filter. The key property of ${\cal F}^{\tytv}$ is to be a $\lambda$-model. This is proved in \cite{aleslusi02}.

\begin{theorem}\label{strong-beta-implies-lm}$\;$\\
The filter model
$\three{\SF}{\cdot}{\interpretation{\;}{\;}{\tytv}}$ is a
$\lambda$-model, in the sense of Hindley-Longo~\cite{hindlong80},
that is:
\begin{enumerate}
\item\label{hl1}
$\interpretation{x}{\rho}{\tytv} = \rho(x)$;
\item\label{hl2}
$\interpretation{\tM\tN}{\rho}{\tytv} =
 \interpretation{\tM}{\rho}{\tytv}\cdot\interpretation{\tN}{\rho}{\tytv}$;
\item\label{hl3}
$\interpretation{\lambda x.\tM}{\rho}{\tytv}\cdot X =
\interpretation{\tM}{\rho[X/x]}{\tytv}$;
\item\label{hl4}
$(\forall x\in\FV(\tM).\; \interpretation{x}{\rho}{\tytv} =
\interpretation{x}{\rho'}{\tytv})\;\Rightarrow\;
\interpretation{\tM}{\rho}{\tytv} =
\interpretation{\tM}{\rho'}{\tytv}$;
\item\label{hl5}
$\interpretation{\lambda x.\tM}{\rho}{\tytv} =
\interpretation{\lambda y.\tM[y/x]}{\rho}{\tytv}$, if
$y\notin\FV(\tM)$;
\item\label{hl6}
$(\forall X\in\SF. \interpretation{\tM}{\rho[X/x]}{\tytv} =
\interpretation{\tN}{\rho[X/x]}{\tytv})\;\Rightarrow\;
\interpretation{\lambda x.\tM}{\rho}{\tytv} =
\interpretation{\lambda x.\tN}{\rho}{\tytv}$.
\end{enumerate}
Moreover it is extensional, that is $\interpretation{\lambda
x.\tM x}{\rho}{\tytv} = \interpretation{\tM}{\rho}{\tytv}$  when
$x\notin\FV(\tM)$.
\end{theorem}

\section{Simple easy terms and filters}\label{section-simple-easy-terms}
In this section we give the main notion of the paper, namely
{\em simple easiness}. A term $\tE$ is simple easy if,
given any eitt $\Sigma^{\tytv}$ and
a type $Z\in\type^\tytv$, we can extend in a conservative
way $\Sigma^{\tytv}$ to $\Sigma^{\tytv'}$, so that
$\interpretation{\tE}{}{\tytv'}=
\FFF{\tytv'} Z\sqcup\interpretation{\tE}{}{\tytv}$.
This allows to build with an uniform
technique filter models in which the
interpretation of $\tE$ is {\em a filter of types induced by a predicate}
(see Definition \ref{nice}).

\begin{definition}\label{filter-scheme}$\;$
\begin{enumerate}
\item
Let be $\sar{\tytv}$ and $\sar{\tytv'}$ two eitts. We say that $\sar{\tytv'}$ is a {\em conservative extension} of $\sar{\tytv}$ (notation
$\sar{\tytv}\sqsubseteq\sar{\tytv'}$) iff
$\con^\tytv\subset\con^{\tytv'}$ and for all
$A,B\in\type^\tytv$,
\[A\leq_\tytv B\;\Leftrightarrow\; A\leq_{\tytv'} B.\]
\item
A {\em pointed\/} eitt is a pair
$(\Sigma^{\tytv}, Z)$ with $Z\in\type^{\tytv}$.
\item A {\em filter scheme\/} is a mapping
$\filterscheme:\mbox{\it PEITT}\rightarrow\mbox{\it EITT}$, such that
for any $(\Sigma^\tytv,Z)$
\[\Sigma^\tytv\sqsubseteq \filterscheme(\Sigma^\tytv,Z),\;\]
where {\it EITT} and
{\it PEITT} denote respectively the classes of
eitts and pointed eitts.
\end{enumerate}
\end{definition}

We now give the central notion of {\em simple easy\/} term.

\begin{definition}\label{simply-easy}$\;$\\
An unsolvable term $\tE$ is {\em simple easy\/} if there
exists a filter scheme $\filterscheme_\tE$ such that for any pointed
eitt $(\Sigma^\tytv, Z)$,
\[\vdash^{ \tytv'} \tE:B \iff \exists
C\in\type^\tytv. C\cap Z\leq_{\tytv'} B\;\&\;\ctas \tE:C,\]
where $\Sigma^{\tytv'} = \filterscheme_\tE(\Sigma^\tytv,Z)$.$\;$
\end{definition}

A first key property of easy terms is showed in  \cite{aleslusi02}.

\begin{theorem}\label{E-prende-Z}$\;$\\
With the same notation of previous definition, we have
$\interpretation{\tE}{}{\tytv'} =
\FFF{\tytv'} Z\sqcup \interpretation{\tE}{}{\tytv}$.
\end{theorem}

The last notion we need is that of filters induced by a predicate.

\begin{definition}\label{nice}$\;$\\\
Let $\p$ be a predicate defined on $\typetytv$ for all $\tytv$.
The $\tytv$-filter induced by $\p$ is the filter defined by:
\[X^{\tytv}_\p=\FFF{\tytv}\set{A\in\typetytv\mid \p(A)}.\]
\end{definition}
%%%%%%%%%%%%%%%%%%%%%%

\begin{theorem}\label{manca}$\;$\\
Let $\tE$ be a simple easy term and $\p$ be as in previous definition. Then there is a filter model in which the interpretation of $\tE$ is the filter induced by $\p$.
\end{theorem}
{\em Proof.}\ \ \ \
Let $\langle \cdot,\cdot\rangle$
denotes any fixed bijection between $\naturali\times\naturali$ and
$\naturali$ such that $\langle r,s\rangle \geq r$.

We will define a denumerable sequence of eitts $\Sigma^{\tytv_{0}}, \ldots,\Sigma^{\tytv_{r}}, \ldots$. For each $r$ we will consider
a fixed enumeration $\langle W^{(r)}_{s}\rangle_{s\in\mbox{I$\!$N}}$ of the set 
$\set{A\in \type^{\tytv_{r}} \mid A\notin \type^{\tytv_{r-1}}\&\p(A)}$ (for $r=0$ the clause
$A\notin \type^{\tytv_{r-1}}$ is vacously true).

We can construct the model as follows.\smallskip

\noindent
{\bf step\/} $0$:\ \ \
take the eitt $\Sigma^{\tytv_{0}}$
whose filter
model is isomorphic to Scott $D_{\infty}$ (see \cite{coppdezahonslong84}):\\
${}$\ \ - $\con^{\tytv_{0}} = \{\tyt,\omega\}$;\\
${}$\ \  - $\tytv_{0} =
    \overline{\tytv}\cup\{\omega\sim\tyt\rightarrow\omega\}$.\smallskip

\noindent
{\bf step\/} $(n+1)$:\ \ \
if $n=\langle r,s\rangle$ we define $\Sigma^{\tytv_{n+1}} =
\filterscheme_{\tE}(\Sigma^{\tytv_{n}}, W^{(r)}_{s})$ (notice that $\Sigma^{\tytv_{n}}\sqsubseteq\Sigma^{\tytv_{n+1}}$);
\smallskip

\noindent
{\bf final step}:\ \ \ take $\Sigma^{\tytv_{*}}=\Sigma(\bigcup_{n} \con^{\tytv_{n}},
\bigcup_{n} \tytv_{n})$.
\smallskip

We prove first that the model ${\cal F}^{\tytv_{*}}$ is non-trivial by
 showing that
$\interpretation{\sf i}{}{\tytv_{*}}\neq
\interpretation{\sf k}{}{\tytv_{*}}$, where ${\sf i}\equiv\lambda x.x$
${\sf k}\equiv\lambda xy.x$. Let $D\equiv(\omega\to\omega)\to(\omega\to\omega)$.
Since
$\vdash^{\tytv_{*}}{\sf i}:D$,
we have that $D\in
\interpretation{\sf i}{}{\tytv_{*}}$.
On the other hand, if it were
$D\in\interpretation{\sf k}{}{\tytv_{*}}$, then it should exists
$n$ such that
$D\in\interpretation{\sf k}{}{\tytv_{n}}$. This would imply
(by applying several times the Generation Theorem)
$\omega\to\omega\leq_{\tytv_{n}} \omega$. Since we have
$\Sigma^{\tytv_{n}}\sqsubseteq \Sigma^{\tytv_{n+1}}$ for any $n$,
we should have $\omega\to\omega\leq_{\tytv_{0}} \omega$.
Since $\omega\sim_{\tytv_{0}}\tyt\to\omega$, we should conclude,
by Theorem \ref{strong-beta-are-beta},
$\tyt\leq_{\tytv_{0}} \omega$, which is a contradiction.
Therefore we cannot have $D\in\interpretation{\sf k}{}{\tytv_{*}}$
and the model ${\cal F}^{\tytv_{*}}$ is non-trivial.

Now we prove that $\interpretation{\tE}{}{\tytv_{*}}=
 \FFF{\tytv_{*}} \set{W^{(r)}_{s} \mid  r,s\in\naturali}$ by showing that 
$\interpretation{\tE}{}{\tytv_{n}}=
 \FFF{\tytv_{n}} \set{W^{(r)}_{s} \mid \langle r,s\rangle< n}$ for all $n$.
The inclusion 
$(\supseteq)$ is immediate by construction. We prove $(\subseteq)$ by induction on $n$.
If $n=0$, then $\interpretation{\tE}{}{\tytv_{0}}=\FFF{\tytv_{0}}\tyt$,
since ${\cal F}^{\tytv_{0}}$ is the Scott $D_{\infty}$ model,
where all unsolvable terms are equated to bottom.
Suppose the thesis true for $n=\langle r_n,s_n\rangle$
and let $B\in\interpretation{\tE}{}{\tytv_{n+1}}$.
Then $\vdash^{\tytv_{n+1}} \tE:B$. This is possible only if
there exists $C\in \type^{\tytv_{n}}$ such that
$C\into W^{(r_n)}_{s_n}\leq_{\tytv_{n+1}} B$ and moreover
$\vdash^{\tytv_{n}} \tE:C$. By induction
we have $C\in\FFF{\tytv_{n}}\set{W^{(r)}_{s} \mid \langle r,s\rangle< n}$,
hence $W^{(r_1)}_{s_1}\into\ldots\into W^{(r_k)}_{s_k}\leq_{\tytv_{n}} C$  for some
$r_1,\ldots, r_k, s_1,\ldots, s_k$ with $\langle r_i,s_i\rangle< n$ $(1\leq i\leq k)$.
We
derive $W^{(r_1)}_{s_1}\into\ldots\into W^{(r_k)}_{s_k}\into W^{(r_n)}_{s_n}\leq_{\tytv_{n+1}} B$, i.e. $B\in  \FFF{\tytv_{n+1}} \set{W^{(r)}_{s} \mid \langle r,s\rangle< n+1}$ .

Finally we show that $A\in\type^{\tytv_{*}}$ and $\p(A)$ iff $A\equiv W^{(r)}_{s}$ for some $r,s$.
If $A\in\type^{\tytv_{*}}$ then there is an $r$ such that $A\in\type^{\tytv_{r}}$ and $A\notin\type^{\tytv_{r-1}}$.   Moreover if $\p(A)$ holds there is $s$ such that 
$A\equiv W^{(r)}_{s}$. The vice versa is immediate. So we can conclude $\interpretation{\tE}{}{\tytv_{*}} = 
 \uparrow \set{A\in \type^{\tytv_{*}} \mid \p(A)}$, i.e. $\interpretation{\tE}{}{\tytv_{*}}=X^{\tytv_*}_\p$.

\section{An application to the consistency of $\lambda$-theories}\label{appl}
We introduce now a $\lambda$-theory whose consistency has been first proved
using a suitable filter model \cite{lusisali02}. We obtain the same model here as a
consequence of Theorem \ref{manca}. Let $\Delta\equiv\lambda x.xx$.

\begin{definition}
The $\lambda$-theory $\cJ$ is
axiomatized by
\[
\gO xx = x;\quad \gO xy = \gO yx;\quad \gO x(\gO yz) = \gO (\gO xy)z.
\]
\end{definition}

It is clear that the previous equations hold if the interpretation of $\gO$ is the join operator on filters.
For using Theorem \ref{manca} we need:
\begin{itemize}
\item $\gO$ to be simple easy;
\item the join operator on filters to be a filter generated by a predicate defined on all types.
\end{itemize}

The first condition is proved in \cite{aleslusi02}. For the second one it is easy to check that the join 
relative to $\SF$ is represented by the filter:
\[\FFF{\tytv_{}}\set{A\to B\to A\into B}.\]
Therefore the required predicate is
\[\p(C)= (C\equiv A\to B\to A\into B).\]

We can conclude:

\begin{theorem}
The $\lambda$-theory $\cJ$ is consistent.
\end{theorem}

Previous result is used in \cite{lusisali02} to show that there exists a sublattice of the lattice of $\lambda$-theories
which satisfies a restricted form of distributivity, called meet
semidistributivity,
and a nontrivial congruence identity (i.e., an identity in the language
of lattices enriched by the relative product of binary relations).

\section{Conclusion}

The notions of simple easy terms and filter models have been successfully applied to show easiness of $\lambda$-terms \cite{aleslusi02}.
The present paper is a first step toward the application of this methodology for proving consistency of $\lambda$-theories. As a side-effect we showed that simple easiness is more general than easiness. The question whether easiness implies simple easiness remains open. An interesting research direction which we plan to follow is the characterization of the $\lambda$-theories whose consistency can be shown using the present approach or some generalizations of it.

\end{document}